\documentclass[12pt]{article}
\usepackage{euscript,amsmath, amssymb, amsfonts}

\input{tcilatex}
\begin{document}

\title{{\Large Oscillator representations for self-adjoint Calogero
Hamiltonians}}
\author{ D.M. Gitman\thanks{%
Institute of Physics, University of Sao Paulo, Brazil; e-mail:
gitman@dfn.if.usp.br}, I.V. Tyutin\thanks{%
Lebedev Physical Institute, Moscow, Russia; e-mail: tyutin@lpi.ru}, and B.L.
Voronov\thanks{%
Lebedev Physical Institute, Moscow, Russia; e-mail: voronov@lpi.ru}}
\date{}
\maketitle

\begin{abstract}
In the article arXiv:0903.5277 [quant-ph], we have presented a
mathematically rigorous quantum-mechanical treatment of a one-dimensional
motion of a particle in the Calogero potential $V\left( x\right) =\alpha
x^{-2}$. In such a way, we have described all possible s.a. operators (s.a.
Hamiltonians) associated with the formal differential expression $\check{H}%
=-d_{x}^{2}+\alpha x^{-2}$ for the Calogero Hamiltonian. Here, we discuss a
new aspect of the problem, the so-called oscillator representation for the
Calogero Hamiltonians. As it is know, operators of the form $\hat{N}=\hat{a}%
^{+}\hat{a}$ and $\hat{A}=\hat{a}\hat{a}^{+}$ are called operators of
oscillator type. Oscillator type operators obey several useful properties in
case if the elementary operator $\hat{a}$ and $\hat{a}^{+}$ are densely
defined. It turns out that some s.a. Calogero Hamiltonians are oscillator
type operators. We describe such Hamiltonians and find the corresponding
mutually adjoint elementary operators.
\end{abstract}

\section{Introduction}

In the article \cite{GitTyV09}, we presented a mathematically rigorous
nonrelativistic quantum-mechanical (QM) treatment of a one-dimensional
motion of a particle in the Calogero potential $V\left( x\right) =\alpha
x^{-2}$ \cite{Calog169}. In this article, we summarily review all essential
mathematical aspects of the one-particle Calogero problem by using a uniform
approach based on the theory of s.a. extensions of symmetric differential
operators, namely, on a method of specifying s.a. ordinary differential
operators associated with s.a. differential expressions by (asymptotic) s.a.
boundary conditions \cite{GitTyV06}. In such a way, we have studied all
possible s.a. operators (s.a. Hamiltonians) associated with the formal
differential expression $\check{H}$ for the Calogero Hamiltonian 
\begin{equation}
\check{H}=-d_{x}^{2}+\alpha x^{-2},\ (d_{x}=d/dx),  \label{6a.2}
\end{equation}%
A complete spectral analysis of all s.a. Hamiltonians was given and the
corresponding complete sets of (generalized) eigenfunctions was found.

In this article, we discuss a new aspect of the problem, the so-called
oscillator representation for the Calogero Hamiltonians. As it is know,
operators of the form $\hat{N}=\hat{a}^{+}\hat{a}$ and $\hat{A}=\hat{a}\hat{a%
}^{+}$ are called operators of oscillator type. Oscillator type operators
obey several useful properties in case if the elementary operator $\hat{a}$
and $\hat{a}^{+}$ are densely defined. Then, in particular, $\left( \hat{a}%
^{+}\right) ^{+}=\hat{a}$, which allows one to call the operators $\hat{a}$
and $\hat{a}^{+}$ mutually adjoint elementary operators. If the operators $%
\hat{a}$ and $\hat{a}^{+}$ are mutually adjoint then the operators $\hat{N}$
and $\hat{A}$\ are s.a. and nonnegative. It turns out that some s.a.
Calogero Hamiltonians discussed in \cite{GitTyV09} are oscillator type
operators. We describe such Hamiltonians and find the corresponding mutually
adjoint elementary operators. One ought to say, that in spite of the fact
that the oscillator representation of s.a. Calogero Hamiltonians is the
principal aim of our consideration, we discuss here preliminarily, as a
particular case of Calogero problem, the oscillator representation of the
free particle Hamiltonian.

\section{General}

Below, we analyze a possibility to represent the Calogero differential
expression $\check{H}$ (\ref{6a.2}) in the form%
\begin{equation}
\check{H}=-d_{x}^{2}+\alpha x^{-2}=\check{b}\check{a},\ \ \check{b}=\check{a}%
^{\dagger },  \label{KC.1}
\end{equation}%
where $\check{a}$ is a differential expressions of finite order and $\check{a%
}^{\dagger }$ its adjoint by Lagrange, see \cite{AkhGl81}.

It is easy to see that $\check{a}$ must be a differential expression of the
form%
\begin{equation}
\check{a}=k(x)d_{x}+\kappa (x),  \label{kc.2}
\end{equation}%
where $k(x)$ and $\kappa (x)$ are some functions of $x.$Then it follows from
(\ref{KC.1}) that $\left\vert k\right\vert ^{2}=1$ and, therefore, $%
k(x)=e^{i\theta (x)}.$ Thus, we can write%
\begin{equation}
\check{a}=e^{i\theta (x)}[d_{x}-h(x)],\ \ \check{b}=\check{a}^{\dagger
}=-[d_{x}+\overline{h(x)}]e^{-i\theta (x)}.  \label{kc.3}
\end{equation}%
The function $\theta (x)$ can be fixed from convenience reasons and will be
set to zero in what follows. Using (\ref{kc.3}) in (\ref{KC.1}), we obtain 
\begin{equation*}
-d_{x}^{2}+\alpha x^{-2}=-d_{x}^{2}+(h-\bar{h})d_{x}+h^{\prime }+\bar{h}h,
\end{equation*}%
which implies that the function $h(x)$ must be real and obey the following
differential equation 
\begin{equation}
h^{\prime }+h^{2}=\alpha x^{-2}.  \label{KC.2}
\end{equation}%
which is a particular case of the Riccati equation.

General solution of eq. (\ref{KC.2}) is%
\begin{equation}
h(x)=\frac{\nu }{x}+\frac{2\varkappa x^{2\varkappa -1}}{c+x^{2\varkappa }},\
\nu =\frac{1}{2}-\varkappa ,\ \varkappa =\sqrt{\alpha +1/4}\ {\LARGE ,}
\label{kc.6}
\end{equation}%
where $c$ is a constant.

The condition that $h(x)$ be real and nonsingular for $x>0$ implies that 
\begin{equation}
\alpha \geq -\frac{1}{4},\ \ \varkappa =\sqrt{\alpha +1/4}\geq 0,\ \ c\geq
0\ .  \label{kc.7}
\end{equation}

We note that in the general case $\check{a}\check{b}\neq -d_{x}^{2}+\alpha
x^{-2}$. However, if $\check{a}$ and $\check{b}$ are still defined by (\ref%
{kc.3}) and we demand%
\begin{equation*}
\check{a}\check{b}=-d_{x}^{2}+\alpha _{1}x^{-2}
\end{equation*}%
with a constant $\alpha _{1}\neq \alpha ,$ then $\theta =0$ and $h$ has to
obey the equation%
\begin{equation}
-h^{\prime }+h^{2}=\alpha _{1}x^{-2}.  \label{kc.13}
\end{equation}%
The function $h(x)$ that obey both equations (\ref{KC.2}) and (\ref{kc.13})
reads%
\begin{equation}
h(x)=\frac{\mu }{x},\ \mu =\sqrt{\frac{\alpha _{1}+\alpha }{2}}=\frac{\alpha
_{1}-\alpha }{2}.  \label{kc.14}
\end{equation}

In fact, we have to study separately three cases:

a)%
\begin{equation}
\alpha=\nu=0,\ 0<c<\infty,\ \varkappa=1/2.  \label{kc.9}
\end{equation}

The case a) corresponds to the free particle differential operation on the
semiaxis.

b) 
\begin{equation}
0<c<\infty ,\ \ \varkappa >0,\ \varkappa \neq \frac{1}{2},  \label{kc.7a}
\end{equation}

c) In the remaining domain of the parameters we have $h\left( x\right) =\mu
x^{-1},$ where%
\begin{align}
c& =0,\ \varkappa >0,\ \ \mu =\nu +2\varkappa ,  \notag \\
c& =\infty ,\ \varkappa >0,\ \ \mu =\nu ,  \notag \\
0& \leq c\leq \infty ,\ \ \varkappa =0,\ \ \mu =\frac{1}{2}.  \label{kc.8}
\end{align}

The case a) is discussed in the next section, whereas the cases b) and c)
are discussed in the last section of the article.

\section{Self-adjoint free-particle Hamiltonians}

For generality, we start with a s.a. differential operation%
\begin{equation}
\check{H}=-d_{x}^{2}+u,  \label{6.3.1}
\end{equation}%
where $u$ is a real constant, having in mind that adding such a constant to
a Hamiltonian can only change the reference point of the energy, which is
not relevant from the physical point of view. Then we are going to try to
find mutually adjoint by Lagrange differential operations $\check{a}$ and $%
\check{b}=\check{a}^{\dag }$ that obey the relation%
\begin{equation}
-d_{x}^{2}+u=\check{b}\check{a}.  \label{6.3.2}
\end{equation}%
If we restrict ourself with finite order differential operations for $\check{%
a}$ and $\check{b},$ then we will see that those can be only differential
operations of the first order that have the form%
\begin{equation}
\check{a}=e^{i\phi (x)}\left[ d_{x}-h(x)\right] ,\ \check{b}=\left[ -d_{x}-%
\bar{h}(x)\right] e^{-i\phi (x)}.  \label{6.3.3}
\end{equation}%
The parameter $\phi (x)$ is not relevant and can be fixed from convenience
considerations. In what follows, we chose $\phi =-\pi /2,$ such that%
\begin{equation}
\check{a}=\check{p}+ih(x),\ \check{b}=\check{p}-i\bar{h}(x).  \label{6.3.3a}
\end{equation}%
Substituting (\ref{6.3.3}) in (\ref{6.3.2}), we obtain%
\begin{equation*}
(h-\bar{h})d_{x}+h^{\prime }+h\bar{h}=u,
\end{equation*}%
which implies that the function $h(x)$ must be real and obey the following
differential equation%
\begin{equation}
h^{\prime }+h^{2}=u.  \label{6.3.4}
\end{equation}

General solution of eq. (\ref{6.3.4}) is described by the following three
families of functions $h_{k}(x)$:

I) $u=s^{2}\geq 0,$ $s\geq 0$

\begin{eqnarray}
h_{1}(x) &=&s\coth [s(x+c_{1})],\ c_{1}\in \mathbb{R},\ |h_{1}(x)|\geq s,
\label{6.3.5} \\
h_{2}(x) &=&s\tanh [s(x+c_{2})],\ c_{2}\in \mathbb{R},\ |h_{2}(x)|\leq s.
\label{6.3.6}
\end{eqnarray}

II) $u=-\sigma ^{2}<0$,$\ \ \sigma >0$

\begin{equation*}
h_{3}(x)=\sigma \cot (\sigma x+c_{3}),\ c_{3}\in \mathbb{R}.
\end{equation*}%
The function $h(x)=(x+c)^{-1}$ from eq. (\ref{kc.6}), for $\alpha =\nu =0,\
0<c<\infty ,\ \varkappa =1/2$, is contained in the family (\ref{6.3.5}) (in
the limit $s\rightarrow 0$ for fixed $c_{1}=c$).

\subsection{Whole real axis}

The solutions $h_{1,3}(x)$ have singularities at finite $x.$ Thus, on the
whole axis, we consider only the solution $h_{2}(x)$ (\ref{6.3.6}), which is
smooth and uniformly bounded on whole axis, \TEXTsymbol{\vert}$h_{2}(x)|\leq
s$, $x\in \mathbb{R}$. Then we are going to construct closed operators
associated with differential expressions $\check{a}_{2}=\check{p}+ih_{2}(x)$,%
$\ \check{b}_{2}=\check{p}-ih_{2}(x)$. First we define the corresponding
initial operators $\hat{a}_{2}$ and $\hat{b}_{2}$ defined on the domain $%
\mathcal{D}(\mathbb{R}),$where they act by their differential expressions
respectively. Then we can write%
\begin{equation}
\hat{a}_{2}=\hat{p}+ih_{2},\ \hat{b}_{2}=\hat{p}-ih_{2},  \label{6.3.10}
\end{equation}%
where the multiplication "operator" $h_{2}$ is a bounded s.a. and defined
everywhere on $L^{2}(\mathbb{R}),$\ and $\hat{p}$ is the initial symmetric
momentum operator of a particle on the whole axis. We remind \cite{AkhGl81,
Naima69} that on the whole axis, the operator $\hat{p}$ is essentially s.a.,
and its unique s.a. extension, let us denote it by $\hat{p}_{{\large %
\epsilon }},$ is its closure, $\hat{p}_{{\large \epsilon }}=\overline{\hat{p}%
}=\hat{p}^{+}$.

It is easy to construct closed operators $\hat{a}_{2}^{+},\hat{b}_{2}^{+},%
\overline{\hat{a}}_{2},$ and $\overline{\hat{b}}_{2}$,%
\begin{align}
& \hat{a}_{2}^{+}=\overline{\hat{p}}-i\hat{h}_{2}=\left\{ 
\begin{array}{l}
D_{a_{2}^{+}}=D_{\bar{p}}=D_{\check{p}}^{\ast}(\mathbb{R}), \\ 
\hat{a}_{2}^{+}\psi=\check{b}_{2}\psi,\ \forall\psi\in D_{\check{p}}^{\ast }(%
\mathbb{R});%
\end{array}
\right.  \notag \\
& \hat{b}_{2}^{+}=\overline{\hat{p}}+i\hat{h}_{2}=\left\{ 
\begin{array}{l}
D_{b_{2}^{+}}=D_{\bar{p}}=D_{\check{p}}^{\ast}(\mathbb{R}), \\ 
\hat{b}_{2}^{+}\psi=\check{a}_{2}\psi,\ \forall\psi\in D_{\check{p}}^{\ast }(%
\mathbb{R});%
\end{array}
\right.  \notag \\
& \overline{\hat{a}}_{2}=(\hat{a}_{2}^{+})^{+}=\hat{b}_{2}^{+},\ \ 
\overline {\hat{b}}_{2}=(\hat{b}_{2}^{+})^{+}=\hat{a}_{2}^{+}.
\label{6.3.11}
\end{align}

We note that does not exist any one closed operators $\hat{g}\neq \overline{%
\hat{a}}_{2}$ with the property $\hat{g}\supseteq\hat{a}_{2}$, $\hat{g}%
^{+}\supseteq\hat{b}_{2}$. Indeed, the chain of inclusions $\hat {a}%
_{2}\subseteq$ $\overline{\hat{a}}_{2}\subseteq\hat{g}\subseteq\hat{b}%
_{2}^{+}=\overline{\hat{a}}_{2}$ implies $\hat{g}=\overline{\hat{a}}_{2}$.

Consider the operator $\hat{a}_{2}^{+}\overline{\hat{a}}_{2}-s^{2}.$
According to the Akhiezer-Glazman theorem, this operator is s.a.. It
coincides with the initial symmetric operator $\widehat{\mathcal{H}}$\ on
the domain $\mathcal{D}(\mathbb{R})$ and, therefore is its s.a. extension.
But it is known that on the whole axis there exists only one such an
extension $\widehat{\mathcal{H}}_{{\large \epsilon }}$\textbf{\ }with the
domain $D_{\mathcal{H}_{{\large \epsilon }}}=D_{\mathcal{\check{H}}}^{\ast
}\left( \mathbb{R}\right) .$ Thus, we obtain the following oscillator
representation on the whole axis%
\begin{equation}
\widehat{\mathcal{H}}_{{\large \epsilon }}+s^{2}=\hat{a}_{2}^{+}\overline{%
\hat{a}}_{2}~.  \label{6.3.12}
\end{equation}

The r.h.s. does not depend on the constant $c_{2}$ which plays role of the
reference point. It is natural in the translation invariant system.

For $s=0$, we have $\overline{\hat{a}}_{2}=\hat{a}_{2}^{+}=\overline{\hat{p}}
$ and the representation (\ref{6.3.12}) is reduced to%
\begin{equation*}
\widehat{\mathcal{H}}_{{\large \epsilon }}=\overline{\hat{p}}^{2}.
\end{equation*}

\subsection{A semiaxis}

Here we will use the solutions $h_{k}(x)$,\ $k=1,2$ that have no
singularities on $\mathbb{R}_{+},$ see (\ref{6.3.5}) and (\ref{6.3.6}). Then
we are going to construct closed operators associated with differential
operations%
\begin{equation}
\check{a}_{k}=\check{p}+ih_{k}(x),\ \check{b}_{k}=\check{p}-ih_{k}(x).
\label{6.3.13}
\end{equation}

First we introduce the corresponding initial operators $\hat{a}_{k}$ and $%
\hat{b}_{k}$ defined on the domain $\mathcal{D}(\mathbb{R}_{+}),$where they
act by the corresponding differential operations (\ref{6.3.13}) respectively,%
\begin{equation}
\hat{a}_{k}=\hat{p}+ih_{k},\ \hat{b}_{k}=\hat{p}-ih_{k}.  \label{6.3.14}
\end{equation}%
Here $h_{k}$ is the multiplication "operators" and defined everywhere on $%
\mathcal{D}(\mathbb{R}_{+})$, \ and $\hat{p}$ is the initial symmetric
momentum operator of a particle on the semiaxis, see sec. 1. We remind that
on the semiaxis, does not exist a s.a. momentum operator, however, there
exist closed operators $\overline{\hat{p}}$ and $\hat{p}^{+}$.

One can easily see that%
\begin{equation}
\left( \xi ,\hat{a}_{k}\psi \right) =\left( \hat{b}_{k}\xi ,\psi \right) ,\
\forall \xi ,\psi \in \mathcal{D}(\mathbb{R}_{+}),  \label{6.3.15}
\end{equation}%
such that the operators $\hat{a}_{k}$ and $\hat{b}_{k}$ are mutually
conjugated on the domain $\mathcal{D}(\mathbb{R}_{+}).$ Since $\hat{a}_{k}$
are densely defined there exist operators $\hat{a}_{k}^{+}$, which act on
their domains $D_{a_{k}^{+}}=D_{\check{b}_{k}}^{\ast }\left( \mathbb{R}%
_{+}\right) $ as $\check{b}_{k}.$ In turn, equation (\ref{6.3.15}) implies
that $\hat{a}_{k}^{+}\supseteq \hat{b}_{k}$, and the operators $\hat{a}%
_{k}^{+}$ are densely defined as well. Then there exist adjoint operators $(%
\hat{a}_{k}^{+})^{+}=\overline{\hat{a}_{k}}$. In the same manner, one can
see that there exist operators $\hat{b}_{k}^{+},$ which acts on its domain $%
D_{b_{k}^{+}}=D_{\check{a}_{k}}^{\ast }\left( \mathbb{R}_{+}\right) $ as $%
\check{a}_{k}.$ In addition $(\hat{b}_{k}^{+})^{+}=\overline{\hat{b}_{k}}$
and the following inclusions hold%
\begin{equation}
\hat{a}_{k}\subset \overline{\hat{a}_{k}}\subseteq \hat{b}_{k}^{+}\ ,\ \ 
\hat{b}_{k}\subset \overline{\hat{b}_{k}}\subseteq \hat{a}_{k}^{+}\ .
\label{6.3.16}
\end{equation}

Consider the equation%
\begin{equation*}
\hat{a}_{k}^{+}\psi_{k}(x)=\eta(x),\ \eta\in L^{2}(\mathbb{R}_{+}),
\end{equation*}
for functions $\psi_{k}(x)\in D_{a_{k}^{+}}$. Its general solution reads%
\begin{align*}
\psi_{k}(x) & =\frac{1}{\tilde{h}_{k}(x)}\left[ B_{k}+i\int_{0}^{x}\tilde{h}%
_{k}(y)\eta(y)dy\right] , \\
\tilde{h}_{1}(x) & =\left\{ 
\begin{array}{c}
\frac{\sinh[s(x+c_{1})]}{\sinh(sc_{1})},\ c_{1}>0 \\ 
\sinh(sx),\ c_{1}=0%
\end{array}
\right. ,\ \tilde{h}_{2}(x)=\frac{\cosh[s(x+c_{2})]}{\cosh(sc_{2})},
\end{align*}
where $B_{1}=0\ $for$\ c_{1}=0.$The latter condition provides $%
\psi_{1}(x)\in L^{2}(\mathbb{R})$. Estimating asymptotic behavior of $%
\psi_{k}$ at $x\rightarrow0$, we obtain%
\begin{equation*}
\psi_{k}(x)=\left\{ 
\begin{array}{l}
B_{k}+O(x^{1/2}),\ c_{1}>0, \\ 
O(x^{1/2}),\ k=1,\ c_{1}=0.%
\end{array}
\right.
\end{equation*}

One can see that there exist a function $\psi _{0k}(x)\in D_{a_{k}^{+}}$,%
\begin{equation*}
\psi _{0k}(x)=\frac{\zeta (x)}{\tilde{h}_{k}(x)},\ \ \check{b}_{k}\psi
_{0k}\left( x\right) =-\frac{i\zeta ^{\prime }(x)}{\tilde{h}_{k}(x)},
\end{equation*}%
where $\zeta (x)\in D_{r}(0,\infty )$ is a fixed smooth functions\footnote{$%
\mathcal{D}_{r}\left( a,b\right) $ is a space of smooth functions with a
support bounded from the right.} equal $1$ in a neighborhood of the point $%
x=0$. This allows one to obtain a convenient representation for the domain $%
D_{a_{k}^{+}}$:%
\begin{align}
D_{a_{k}^{+}}& =\left\{ 
\begin{array}{l}
B_{k}\psi _{0k}+\mathbb{T}_{_{k}},\ c_{1}>0 \\ 
\mathbb{T}_{_{1}},\ k=1,\ c_{1}=0%
\end{array}%
\right. ,  \notag \\
\mathbb{T}_{_{k}}& =\{\psi :\psi \in D_{\check{a}_{k}^{\dag }}^{\ast }\left( 
\mathbb{R}_{+}\right) ,\ \psi =O(x^{1/2}),\ x\rightarrow 0,  \label{5.2.2.2}
\end{align}%
on which $\hat{a}_{k}^{+}$ acts as $\check{a}_{k}^{\dag }=\check{b}_{k}.$

Relations (\ref{5.2.2.2}) represent here the von Neumann formula.

Similarly, we can demonstrate that functions $\psi_{k}\in D_{b_{k}^{+}}$
have the following representation%
\begin{equation}
\psi_{k}(x)=\tilde{h}_{k}(x)\left[ A_{k}+i\int_{x_{0}}^{x}\frac{\eta (y)}{%
\tilde{h}_{k}(y)}dy\right] ,  \label{5.2.2.3}
\end{equation}
where $x_{0}=0,\ c_{1}>0,$ and $x_{0}>0,\ k=1,\ c_{1}=0.$ Estimating
asymptotic behavior of $\psi_{k}\in D_{b_{k}^{+}}$ at $x\rightarrow0$, we
obtain%
\begin{equation*}
\psi_{k}(x)=\left\{ 
\begin{array}{c}
A_{k}+O(x^{1/2}),\ c_{1}>0, \\ 
O(x^{1/2}),\ k=1,\ c_{1}=0.%
\end{array}
\right.
\end{equation*}
One can see that there exist a function $\xi_{0k}(x)\in D_{b_{k}^{+}}\ ,$%
\begin{equation*}
\xi_{0k}\left( x\right) =\zeta(x)\tilde{h}_{k}(x),\ \check{a}_{k}\xi
_{0k}\left( x\right) =-i\zeta^{\prime}(x)\tilde{h}_{k}(x).
\end{equation*}
This allows one to obtain a convenient representation for the domain $%
D_{b_{k}^{+}}$:%
\begin{align}
& D_{b_{k}^{+}}=\left\{ 
\begin{array}{c}
A_{k}\xi_{0k}+\Upsilon_{k},\ c_{1}>0 \\ 
\Upsilon_{1}\ ,\ \ k=1,\ c_{1}=0%
\end{array}
\right. ,  \label{5.2.2.3a} \\
& \Upsilon_{k}=\{\psi:\psi\in D_{\check{b}_{k}^{\dag}}^{\ast}\left( \mathbb{R%
}_{+}\right) ,\ \psi=O(x^{1/2}),\ x\rightarrow0,  \notag
\end{align}
on which $\hat{b}_{k}^{+}$ acts as $\check{b}_{k}^{\dag}=\check{a}_{k}$.

We note that all the functions from the domains $D_{a_{k}^{+}}$, $\mathbb{T}%
_{_{k}}$, $D_{b_{k}^{+}}$, and $\Upsilon _{k}$ vanish as $x\rightarrow
\infty $. Indeed, the functions $h_{k}(x)$ are bounded at the infinity, that
is why the conditions $\psi ,\check{a}_{k}\psi \in L^{2}(a,\infty )$ or $%
\psi ,\check{b}_{k}\psi \in L^{2}(a,\infty )$ are equivalent to the ones $%
\psi ,\psi ^{\prime }\in L^{2}(a,\infty )$. This allows us to prove the
assertion.

Consider the operators $\overline{\hat{a}_{k}}=(\hat{a}_{k}^{+})^{+}.$ The
inclusion $\overline{\hat{a}_{k}}\subseteq \hat{b}_{k}^{+}$ implies $D_{\bar{%
a}_{k}}\subseteq D_{b_{k}^{+}}$ and that $\overline{\hat{a}_{k}}$ act as $%
\check{a}_{k}$ on their domains. The defining equation for $\overline{\hat{a}%
_{k}}$, which is%
\begin{align*}
& (\psi ,\overline{\hat{a}_{k}}\xi )-(\hat{a}_{k}^{+}\psi ,\xi )=\overline{%
B_{k}}A_{k}\left[ (\psi _{0k},\check{a}_{k}\xi _{0k})-(\check{b}_{k}\psi
_{0k},\xi _{0k})\right] \\
& =2i\overline{B_{k}}A_{k}=0,\ \forall \xi \in D_{\bar{a}_{k}},\ \forall
\psi \in D_{a_{k}^{+}},\ \forall B_{k}\ ,
\end{align*}%
implies $A_{k}=0$. Thus, 
\begin{equation}
\overline{\hat{a}_{k}}:\left\{ 
\begin{array}{l}
D_{\bar{a}_{k}}=\Upsilon _{k}\ , \\ 
\overline{\hat{a}_{k}}\psi =\check{a}_{k}\psi ,\ \ \forall \psi \in D_{\bar{a%
}_{k}}%
\end{array}%
\right. ,  \label{6.3.20}
\end{equation}%
in particular, $D_{\bar{a}_{1}}=D_{b_{1}^{+}},\ \ \overline{\hat{a}_{1}}=%
\hat{b}_{1}^{+}\ $for$\ c_{1}=0.$

Consider the operators $(\hat{b}_{k}^{+})^{+}=\overline{\hat{b}_{k}}\ .$As
before, we can see that%
\begin{equation}
\overline{\hat{b}_{k}}:\left\{ 
\begin{array}{l}
D_{\bar{b}_{k}}=\mathbb{T}_{_{k}}\ , \\ 
\overline{\hat{b}_{k}}\psi=\check{b}_{k}\psi,\ \forall\psi\in D_{\bar{b}%
_{k}},%
\end{array}
\right.  \label{6.3.21}
\end{equation}
in particular, $D_{\bar{b}_{1}}=D_{a_{1}^{+}},\ \ \overline{\hat{b}_{k}}=%
\hat{a}_{1}^{+}\ $for$\ c_{1}=0.$

Taking into account the fact that all the functions $h_{k}(x)$, with the
exception of the function $h_{1}(x)$ for $c_{1}=0$, are bounded on $\mathbb{R%
}_{+}$, we can see that the constructed closed operators are expressed via
momentum operators on the semiaxis as follows:%
\begin{align}
\hat{a}_{k}^{+}& =\hat{p}^{+}-ih_{k},\ \hat{b}_{k}^{+}=\hat{p}^{+}+ih_{k}, 
\notag \\
\overline{\hat{a}}_{k}& =\overline{\hat{p}}+ih_{k},\ \overline{\hat{b}}_{k}=%
\overline{\hat{p}}-ih_{k},\ \ c_{1}>0.  \label{6.3.22}
\end{align}%
For $s=0$ and $c_{1}=\infty $, we find%
\begin{equation*}
\hat{a}_{k}^{+}=\hat{b}_{k}^{+}=\hat{p}^{+},\ \overline{\hat{a}}_{k}=%
\overline{\hat{b}}_{k}=\overline{\hat{p}},\ s=0,\ c_{1}=\infty .
\end{equation*}

One can see that if:

a) a closed operator $\hat{g}$ obeys the properties $\hat{g}\supset 
\overline{\hat{a}}_{k}$, $\hat{g}^{+}\supset \overline{\hat{b}}_{k},$ then
either $\hat{g}=\overline{\hat{a}}_{k}$ or $\hat{g}=\hat{b}_{k}^{+}$;

b) a closed operator $\hat{g}$ obeys the properties $\hat{g}\supset \hat{b}%
_{k}$, $\hat{g}^{+}\supset \hat{a}_{k},$ then either $\hat{g}=\overline{\hat{%
b}_{k}}$ or $\hat{g}=\hat{a}_{k}^{+}$.

Indeed, let $\hat{g}$ obeys properties from a). Then we have the inclusions $%
\hat{a}_{k}\subset \overline{\hat{a}_{k}}\subseteq \hat{g}\subseteq \hat{b}%
_{k}^{+}.$ They imply that $D_{g}\subseteq D_{b_{k}^{+}}$ and the operator $%
\hat{g}$ acts as $\check{a}_{k}$ on $D_{g}$. Then a) follows from the
structure (\ref{5.2.2.3a}) of the domain $D_{b_{k}^{+}}$. Similarly, b)
follows from the structure (\ref{5.2.2.2}) of the domain $D_{a_{k}^{+}}$.

Consider the operators 
\begin{equation*}
\hat{N}_{k}=\hat{a}_{k}^{+}\hat{a}_{k}-s^{2},\ \ \hat{A}_{k}=\hat{b}_{k}\hat{%
b}_{k}^{+}-s^{2},\ \ s\geq 0.
\end{equation*}%
According to the Akhiezer-Glazman theorem \cite{AkhGl81}, all these
operators are s.a. and coincide with the initial symmetric operator $%
\widehat{\mathcal{H}}$ on the domain $\mathcal{D}(\mathbb{R}_{+}).$ That is
why they are some s.a. extensions $\widehat{\mathcal{H}}_{\mathfrak{e}}$ of $%
\widehat{\mathcal{H}}.$ To identify these extensions, we need to identify
only the corresponding domains, since all the operators $\hat{N}_{k}$, $\hat{%
A}_{k}$, and $\widehat{\mathcal{H}}_{\mathfrak{e}}$ act as $\mathcal{\check{H%
}}$ on their domains.

We do not consider such a procedure here (it will be published later). We
only note that each s.a. operator $\widehat{\mathcal{H}}_{\mathfrak{e}}$ has
a generalized (not unique) oscillator representation.

\subsection{A finite interval}

In this case we have a number of closed operators associated with
differential expressions $\check{a}$ and $\check{b}$, since we can use
functions $h_{3}$, and due to new possibilities for constructing closed
extensions. Here, we will not discuss all the detail, restricting ourselves
by an example.

Let us consider a free particle on the interval $(0,l)$ of the real axes. A
domain of the initial symmetric operators $\hat{p}$ and $\widehat{\mathcal{H}%
}$, we define as $\mathcal{D}(0,l),$ on such a domain they act as $\check{p}%
= $ $-id_{x}$ and $\mathcal{\check{H}=}-d_{x}^{2}$ respectively. The
corresponding adjoint operators $\hat{p}^{+}$ and $\widehat{\mathcal{H}}^{+}$
have the same action on the domains $D_{\check{p}}^{\ast }\left( 0,l\right) $
and $D_{\mathcal{\check{H}}}^{\ast }\left( 0,l\right) $ respectively, where $%
D_{\check{p}}^{\ast }\left( 0,l\right) $ is a space of absolutely continuous
(a.c.) functions on $[0,l],$ and $D_{\mathcal{\check{H}}}^{\ast }\left(
0,l\right) $ is a space of a.c. functions together with their first
derivatives on $[0,l]$. The closure $\overline{\hat{p}}$ of the operator $%
\hat{p}$ is defined on the domain $D_{\bar{p}}=\{\psi :\psi \in D_{\check{p}%
}^{\ast }\left( 0,l\right) ,\ \psi (0)=\psi (l)=0\}$, where it acts as $%
\check{p}$.

Consider the operator $\hat{p}^{+}\overline{\hat{p}}.$ Let us find functions 
$\psi \in D_{p^{+}\overline{p}}$ . First of all, $\psi \in D_{\bar{p}}$. In
addition, the functions $\overline{\hat{p}}\psi \left( x\right) $ must
belong to $D_{p^{+}}$. These conditions allow one to find the domain $%
D_{p^{+}\overline{p}}$,%
\begin{equation*}
D_{p^{+}\overline{p}}=\{\psi :\psi ,\psi ^{\prime }\ \mathrm{a.c.\ on}\
[0,l],\ \psi (0)=\psi (l)=0\}.
\end{equation*}%
We note that $D_{p^{+}\overline{p}}\in D_{\mathcal{\check{H}}}^{\ast }\left(
0,l\right) $. According to the Akhiezer-Glazman theorem \cite{AkhGl81}, the
operator $\hat{p}^{+}\overline{\hat{p}}$ is s.a. nonnegative operator. On
the domain $\mathcal{D}(0,l)\subset D_{p^{+}\overline{p}}$ such an operator
coincides with the symmetric operator $\widehat{\mathcal{H}}$ and, therefore
is one of its s.a. extensions $\widehat{\mathcal{H}}_{\mathfrak{e}}$. Thus,
we obtain $\widehat{\mathcal{H}}_{\mathfrak{e}}=\hat{p}^{+}\overline{\hat{p}}%
,$ where $D_{\mathcal{H}_{\mathfrak{e}}}=\{\psi :\psi \in D_{\mathcal{\check{%
H}}}^{\ast }\left( 0,l\right) ,\ \psi (0)=\psi (l)=0\}$. From the physical
point of view, the latter operator represents a Hamiltonian of the free
particle in the infinite rectangular potential well.

\section{Self-adjoint Calogero Hamiltonians}

\subsection{The case b)}

Here%
\begin{equation}
\check{a}=d_{x}-h(x),\ \check{b}=-d_{x}-h(x)\ ,\ h(x)=\frac{\nu }{x}+\frac{%
2\varkappa x^{2\varkappa -1}}{c+x^{2\varkappa }}.  \label{kc.15}
\end{equation}%
Let us introduce initial operators $\hat{a}$ and $\hat{b}$ defined on the
domain $\mathcal{D}(\mathbb{R}_{+}),$where they act by their differential
expressions respectively. The operators have the following properties%
\begin{align}
& \left( \psi ,\hat{a}\xi \right) =\left( \hat{b}\psi ,\xi \right)
,\;\forall \psi ,\xi \in \mathcal{D}(\mathbb{R}_{+})~,  \label{KC.3} \\
& \hat{H}=\hat{b}\hat{a}\ ,  \label{KC.4} \\
& \left( \xi ,\hat{H}\xi \right) =\left( \hat{a}\xi ,\hat{a}\xi \right)
>0,\;\forall \xi \in \mathcal{D}(\mathbb{R}_{+}).  \label{KC.5}
\end{align}%
The property (\ref{KC.3}) allows us to treat the operators $\hat{a}$ and $%
\hat{b}$ as mutually adjoint on the domain $\mathcal{D}(\mathbb{R}_{+})$.
Eqs. (\ref{KC.4}) and (\ref{KC.5}) imply that the initial symmetric operator 
$\hat{H}$ for $\alpha \geq -1/4$ is nonnegative.

Since the operator $\hat{a}$ is densely defined, there exists its adjoint $%
\hat{a}^{+}$ which is a closed operator. The defining equation for $\hat {a}%
^{+}$ reads%
\begin{equation}
(\eta,\xi)=(\psi,\hat{a}\xi),\ \ \forall\xi\in\mathcal{D}(\mathbb{R}_{+}),\
\psi\in D_{a^{+}},\ \ \eta=\hat{a}^{+}\psi.  \label{KC.6}
\end{equation}
Due to the property (\ref{KC.3}), one can see that eq. (\ref{KC.6}) has
solutions of the form%
\begin{equation}
\psi=\xi,\ \ \eta=\hat{b}\xi,\ \ \xi\in\mathcal{D}(\mathbb{R}_{+}),
\label{kc.17}
\end{equation}
which implies $\hat{b}\subseteq\hat{a}^{+}$. Therefore $\hat{a}^{+}$ is
densely defined and has an adjoint operator $(\hat{a}^{+})^{+}$ that is
obviously closed. In this case, the operator $\hat{a}$\ admits a closure $%
\overline{\hat{a}}=(\hat{a}^{+})^{+}$; in addition $\overline{\hat{a}}\supset%
\hat{a}$. In the same manner, we find that $\hat{b}^{+}\supset\hat{a}$ and
that the operator $\overline{\hat{b}}=(\hat{b}^{+})^{+}\supset\hat{b}$ does
exist. In addition, it is clear that if an operator $\hat{A}$ is closed and $%
\hat{A}\supseteq\hat{B}$, then $\hat{A}\supseteq\overline{\hat{B}}\supseteq%
\hat{B}$. This allows us easily to generalize the previous inclusions%
\begin{equation}
\hat{a}^{+}\supseteq\overline{\hat{b}}\supset\hat{b},\;\hat{b}^{+}\supseteq%
\overline{\hat{a}}\supset\hat{a}\ .  \label{KC.7}
\end{equation}

Below, we describe domains of the above-introduced operators $\hat{a}^{+},\ 
\hat{b}^{+}\ ,\overline{\hat{a}},$ and$\ \overline{\hat{b}}$\ .\ In doing
this, we are going to follow \cite{AkhGl81,Naima69} and to use the fact that 
$\psi (x)\rightarrow 0$ as $x\rightarrow \infty $ for $\psi (x)\in D_{\check{%
a}^{\dagger }}^{\ast }\left( \mathbb{R}_{+}\right) .$ The latter is due to
the fact \ that the function $h\left( x\right) $ tends to zero as $%
x\rightarrow \infty $.

a) The operator $\hat{a}^{+}$ is defined on the natural domain $D_{\check{a}%
^{\dagger }}^{\ast }\left( \mathbb{R}_{+}\right) ,$ i.e., $D_{a^{+}}=D_{%
\check{a}^{\dagger }}^{\ast }\left( \mathbb{R}_{+}\right) ,$ see \cite%
{GitTyV06,AkhGl81,Naima69}, where it acts as $\check{b}=\check{a}^{\dagger }$%
.\ The functions $\psi \in D_{a^{+}}$ can be represented as%
\begin{equation}
\psi (x)=\frac{x^{-\nu }}{c+x^{2\varkappa }}\left[ A+\int_{x_{0}}^{x}dyy^{%
\nu }(c+y^{2\varkappa })\eta (y)\right] ,\;\eta \in L^{2}(\mathbb{R}_{+}),
\label{kc.19}
\end{equation}%
where $x_{0}=0\ $for$\ \varkappa <1;$\ $x_{0}>0\ $for$\ \varkappa \geq 1$,
and $A$ is a constant. In fact, (\ref{kc.19}) is the general solution of the
equation $\check{a}^{\dagger }\psi =\eta \in L^{2}(\mathbb{R}_{+}).$ This
solution has the following asymptotic behavior:%
\begin{align}
& \psi (x)=A\frac{x^{-\nu }}{c+x^{2\varkappa }}+\left\{ 
\begin{array}{c}
O(x^{1/2}),\ \varkappa \neq 1, \\ 
O(x^{1/2}\ln ^{1/2}x),\ \varkappa =1,%
\end{array}%
\right. ,\;x\rightarrow 0.  \notag \\
& \psi (x)\rightarrow 0,\;x\rightarrow \infty ,  \label{kc.20a}
\end{align}

We note that the domain $D_{a^{+}}$ contains a function $\psi _{0}$,%
\begin{equation}
\psi _{0}(x)=\frac{x^{-\nu }}{c+x^{2\varkappa }}\zeta \left( x\right) ,\ \ 
\check{a}^{\dagger }\psi _{0}(x)=-\frac{x^{-\nu }}{c+x^{2\varkappa }}\zeta
^{\prime }\left( x\right) ,  \label{kc.21}
\end{equation}%
where $\zeta (x)\in \mathcal{D}_{r}(\mathbb{R}_{+})$ is one (but a fixed) of
smooth function equal $1$ in a neighborhood of the point $x=0$. Then the
domain $D_{a^{+}}$ can be represented as $D_{a^{+}}=\tilde{D}_{\check{a}%
^{\dagger }}^{\ast }\left( \mathbb{R}_{+}\right) +A\psi _{0},$ where $\tilde{%
D}_{\check{a}^{\dagger }}^{\ast }\left( \mathbb{R}_{+}\right) $ is
restriction of $D_{\check{a}^{\dagger }}^{\ast }\left( \mathbb{R}_{+}\right) 
$ to functions $\psi $ with the following asymptotic behavior%
\begin{align}
& \psi =\left\{ 
\begin{array}{l}
O(x^{1/2}),\ \ \varkappa \neq 1, \\ 
O(x^{1/2}\sqrt{\ln x}),\ \ \varkappa =1,%
\end{array}%
\right. ,\;x\rightarrow 0;  \notag \\
& \psi \rightarrow 0,\ x\rightarrow \infty ~.  \label{kc.23}
\end{align}

For $\varkappa \geq 1$ all the function from $D_{\check{a}^{\dagger }}^{\ast
}\left( \mathbb{R}_{+}\right) $ have the behavior (\ref{kc.23}), therefore
the functions $A\chi _{0}\left( x\right) $ belong to $\tilde{D}_{\check{a}%
^{\dagger }}^{\ast }\left( \mathbb{R}_{+}\right) ,$ and in such a case $%
D_{a^{+}}=\tilde{D}_{\check{a}^{\dagger }}^{\ast }\left( \mathbb{R}%
_{+}\right) =D_{\check{a}^{\dagger }}^{\ast }\left( \mathbb{R}_{+}\right) $%
~. Thus,%
\begin{equation}
D_{a^{+}}=\left\{ 
\begin{array}{l}
\tilde{D}_{\check{a}^{\dagger }}^{\ast }\left( \mathbb{R}_{+}\right) +A\psi
_{0},\ 0<\varkappa <1, \\ 
\tilde{D}_{\check{a}^{\dagger }}^{\ast }\left( \mathbb{R}_{+}\right) ,\ \
\varkappa \geq 1.%
\end{array}%
\right.  \label{kc.23b}
\end{equation}

b) The operator $\hat{b}^{+}$ is defined on the natural domain $D_{\check{a}%
}^{\ast }\left( \mathbb{R}_{+}\right) ,$ i.e., $D_{b^{+}}=D_{\check{a}%
}^{\ast }\left( \mathbb{R}_{+}\right) ,$ where it acts as $\check{a}=\check{b%
}^{\dagger }$. Functions $\chi \in D_{b^{+}}$ can be represented as%
\begin{align}
& \chi (x)=x^{\nu }(c+x^{2\varkappa })\left[ B+\int_{0}^{x}dyy^{-\nu
}(c+y^{2\varkappa })^{-1}\eta (y)\right] ,\;\eta \in L^{2}(\mathbb{R}_{+}), 
\notag \\
& B=0\ \ \mathrm{for\ }\varkappa \geq 1,  \label{KC.7a}
\end{align}%
where $B$ and $c$ are some constants. The constant $B$ has to be zero for$%
\mathrm{\ }\varkappa \geq 1$ due to the condition $\chi (x)\in L^{2}(\mathbb{%
R}_{+}).$

The functions (\ref{KC.7a}) have the following asymptotic behavior%
\begin{align}
& \chi(x)\rightarrow0,\;x\rightarrow\infty\ ,  \notag \\
& \chi(x)=\left\{ 
\begin{array}{l}
Bx^{\nu}(c+x^{2\varkappa})+O(x^{1/2}),\;x\rightarrow0,\ \ 0\leq\varkappa <1\
, \\ 
O(x^{1/2}),\;x\rightarrow0,\ \ \varkappa\geq1\ .%
\end{array}
\right.  \label{kc.24b}
\end{align}

We note that the domain $D_{b^{+}}$ contains a function $\chi _{0}(x)$,%
\begin{equation*}
\chi _{0}(x)=x^{\nu }(c+x^{2\varkappa })\zeta (x),\ \check{b}^{\dagger }\chi
_{0}(x)=x^{\nu }(c+x^{2\varkappa })\zeta ^{\prime }(x).
\end{equation*}%
Then the domain $D_{b^{+}}$ can be represented in the form%
\begin{equation}
D_{b^{+}}=\left\{ 
\begin{array}{l}
\tilde{D}_{\check{a}}^{\ast }\left( \mathbb{R}_{+}\right) +B\chi _{0},\ \
0<\varkappa <1, \\ 
\tilde{D}_{\check{a}}^{\ast }\left( \mathbb{R}_{+}\right) ,\ \ \varkappa
\geq 1,%
\end{array}%
\right.  \label{kc.24a}
\end{equation}%
where $\tilde{D}_{\check{a}}^{\ast }\left( \mathbb{R}_{+}\right) $ is
restriction of $D_{\check{a}}^{\ast }\left( \mathbb{R}_{+}\right) $ to
functions $\chi $ with the following asymptotic behavior 
\begin{equation}
\chi (x)=O(x^{1/2}),\;x\rightarrow 0;\ \ \chi (x)\rightarrow 0,\ \
x\rightarrow \infty \ .  \label{kc.27a}
\end{equation}

c) The operator $\overline{\hat{a}}$ we construct as $\overline{\hat{a}}%
=\left( \hat{a}^{+}\right) ^{+}.$ The defining equation for $\left( \hat{a}%
^{+}\right) ^{+}$ is

\begin{equation}
(\eta,\psi)=(\chi,\hat{a}^{+}\psi),\ \ \forall\psi\in D_{a^{+}},\ \chi\in D_{%
\bar{a}},\ \ \eta=\left( \hat{a}^{+}\right) ^{+}\chi.  \label{kc.26}
\end{equation}
As was established above, $\overline{\hat{a}}\subseteq\hat{b}^{+}.$ This
means that on its domain $D_{\bar{a}}$, the operator $\overline{\hat{a}}$
acts by the differential operation $\check{a}=\check{b}^{\dagger}$, whereas $%
D_{\bar{a}}\subset D_{b^{+}}=\tilde{D}_{\check{a}}^{\ast}\left( \mathbb{R}%
_{+}\right) +B\chi_{0}\ .$ Taking into account (\ref{kc.19}) and (\ref{KC.7a}%
), the asymptotic behavior of the corresponding functions, and, integrating
by parts in (\ref{kc.26}), we find 
\begin{align}
& (\chi,\hat{a}^{+}\psi)-(\eta,\psi)=(\chi,\check{b}\psi)-(\check{a}\chi
,\psi)  \notag \\
& =A\bar{B}[(\chi_{0},\check{b}\psi_{0})-(\check{a}\chi_{0},\psi_{0})]=2A%
\bar{B}=0,\ \forall A\ .  \label{kc.28}
\end{align}
The latter results implies $B=0,$which means in turn,%
\begin{equation}
D_{\bar{a}}=\tilde{D}_{\check{a}}^{\ast}\left( \mathbb{R}_{+}\right) \ .
\label{kc.27}
\end{equation}

We note that $\hat{b}^{+}=\overline{\hat{a}}$ for $\varkappa\geq1$, and $%
\hat{b}^{+}\supset\overline{\hat{a}}$ for $0<\varkappa<1$.

d) Similar consideration allow us to find that the operator $\overline{\hat {%
b}}$ is defined on the domain $D_{\bar{\gamma}}=\tilde{D}_{\check{b}}^{\ast
}\left( \mathbb{R}_{+}\right) ,$ where it acts as $\check{b}.$ We note that $%
\hat{a}^{+}=\overline{\hat{b}}$ for $\varkappa\geq1$, and $\hat{a}^{+}\supset%
\overline{\hat{b}}$ for $0<\varkappa<1$.

Thus, in the case under consideration, we have demonstrated that a mutually
conjugated by Lagrange pair of differential operations $\check{a}$ and $%
\check{b}\ \left( \check{a}=\check{b}^{\dagger },\ \check{b}=\check{a}%
^{\dagger }\right) $ generates two mutually conjugated closed operator pairs 
$\overline{\hat{a}}$ and $\hat{a}^{+}\ \left( \overline{\hat{a}}=\left( \hat{%
a}^{+}\right) ^{+}\text{,}\ \hat{a}^{+}=\left( \overline{\hat{a}}\right)
^{+}\right) $ and $\overline{\hat{b}}$ and $\hat{b}^{+}\ \left( \overline{%
\hat{b}}=\left( \hat{b}^{+}\right) ^{+}\text{,}\ \hat{b}^{+}=\left( 
\overline{\hat{b}}\right) ^{+}\right) .$ On their domains the operators $%
\overline{\hat{a}}$ and $\hat{b}^{+}$ act as $\check{a},$ and the operators $%
\hat{a}^{+}$ and $\overline{\hat{b}}$ act as $\check{b}.$

One can see that does not exist any mutually conjugated closed operator
pair, let say $\hat{g}$ and $\hat{g}^{+}$, which obeys the natural property%
\begin{equation}
\hat{g}\varphi=\hat{a}\varphi,\ \hat{g}^{+}\varphi=\hat{b}\varphi ,\
\forall\varphi\in\mathcal{D}(\mathbb{R}_{+})  \label{KC.8c}
\end{equation}
and is different from the two above described pairs.

Indeed, for any closed operator $\hat{g}$ that obeys (\ref{KC.8c}), we have%
\begin{equation*}
\hat{a}\subset \overline{\hat{a}}\subseteq \hat{g}\subseteq \hat{b}^{+},\ \
D_{a}\subset D_{\bar{a}}\subseteq D_{g}\subseteq D_{b^{+}}
\end{equation*}%
and on its domain $\hat{g}$ acts as $\check{a}.$ The structure (\ref{kc.24a}%
) of the domain $D_{b^{+}}$ implies that $D_{g}$ coincides either with $D_{%
\bar{a}}$ or with $D_{b^{+}},$ which means that $\hat{g}=\overline{\hat{a}}$
or $\hat{g}=\hat{b}^{+}$. A similar assertion: $\hat{g}=\overline{\hat{b}}$
or $\hat{g}=\hat{a}^{+}$, holds for the conjugated pair $\hat{g}_{1}$ and $%
\hat{g}_{1}^{+}$, which obeys the property%
\begin{equation*}
\hat{g}_{1}\varphi =\hat{b}\varphi ,\ \hat{g}_{1}^{+}\varphi =\hat{a}\varphi
,\ \ \forall \varphi \in \mathcal{D}(\mathbb{R}_{+}).
\end{equation*}

Now we are in position to construct oscillator representations for some s.a.
Calogero Hamiltonians.

According to the Akhiezer-Glazman theorem, see \cite{AkhGl81}, the operators 
$\hat{A}=\hat{a}^{+}\overline{\hat{a}}$ and $\hat{B}=\overline{\hat{b}}\hat{b%
}^{+}$ are nonnegative s.a. operators. Since $\hat{A}\varphi =\hat{B}\varphi
=\hat{b}\hat{a}\varphi ,$ $\forall \varphi \in \mathcal{D}(\mathbb{R}_{+})$,
both $\hat{A}$ and $\hat{B}$ coincide on the domain $\mathcal{D}(\mathbb{R}%
_{+})$ with the initial symmetric Calogero operator $\hat{H}$. This means
that both $\hat{A}$ and $\hat{B}$ are s.a. extensions of $\hat{H}$ that act
on their domains as $\check{H}=\hat{b}\hat{a}.$ This agrees with the fact
that any s.a. extension of $\hat{H},$ acts as $\check{H}$; all them differ
only by their domains. That is why, we can identify the operators $\hat{A}$
and $\hat{B}$ with certain s.a. Calogero Hamiltonians identifying the
corresponding domains.

Let $\varkappa\geq1$.

In such a case $\hat{A}=\hat{B}=\hat{a}^{+}\overline{\hat{a}}$. On the other
side, in this case there exists only one s.a. Calogero Hamiltonian $\hat{H}%
_{1},$ see \cite{GitTyV09}. That is why we can immediately conclude that%
\begin{equation}
\hat{H}_{1}=\hat{a}^{+}\overline{\hat{a}},\ \ \varkappa \geq 1.
\label{kc.30}
\end{equation}

Let $0<\varkappa<1$, $\varkappa\neq1/2$.

Consider the operator $\hat{A}=\hat{a}^{+}\overline{\hat{a}}$. Its domain $%
D_{A}$ consists of functions $\chi\in D_{\bar{a}}$ that must obey the
relation 
\begin{equation}
\overline{\hat{a}}\chi\in D_{a^{+}}.  \label{KC.9}
\end{equation}

The domain $D_{\bar{a}}$ (\ref{kc.27}) consists of functions from $D_{\check{%
a}}^{\ast }\left( \mathbb{R}_{+}\right) $ which have the asymptotic behavior
(\ref{kc.27a}), in particular, $\chi =O(x^{1/2})$ as $x\rightarrow 0$. The
additional condition (\ref{KC.9}) can only reinforce the asymptotic.
Therefore, the functions from $D_{A}$ are those from $D_{\bar{a}}$ with the
property: they tend to zero not weaker than $x^{1/2}$ as $x\rightarrow 0$.
As follows from \cite{GitTyV09}, there exists only one s.a. Calogero
Hamiltonian with such a domain. It is $\hat{H}_{2,0}.$ Thus,%
\begin{equation}
\hat{H}_{2,0}=\hat{a}^{+}\overline{\hat{a}},\ \ 0<\varkappa <1,\ \varkappa
\neq 1/2\ .  \label{kc.31}
\end{equation}

Consider the operator $\hat{B}=\overline{\hat{b}}\hat{b}^{+}$. Its domain $%
D_{B}$ consists of functions $\chi \in D_{b^{+}}$ that must obey the
relation 
\begin{equation*}
\hat{b}^{+}\chi \in D_{\bar{\gamma}}.
\end{equation*}%
The functions $\chi \in D_{b^{+}}$ have the form (\ref{KC.7a}), which implies%
\footnote{%
In fact, representation (\ref{KC.7a}) is a consequence of (\ref{KC.10}).}%
\begin{equation}
\hat{b}^{+}\chi =\check{a}\chi =\eta \in L^{2}\left( \mathbb{R}_{+}\right) .
\label{KC.10}
\end{equation}%
It follows from (\ref{KC.10}) that $\eta $ are functions from $D_{\bar{\gamma%
}}$ that have the following asymptotic behavior%
\begin{equation}
\ \eta (x)=O(x^{1/2}),\ \ x\rightarrow 0.  \label{KC.11}
\end{equation}%
Estimating by the help of Cauchy-Bunyakovskii the integral summand in (\ref%
{KC.7a})$\ $at $x\rightarrow 0,$ with $\eta (x)$ obeying (\ref{KC.11}), we
find $\left. \chi \left( x\right) \right\vert _{B=0}=O(x^{3/2}),\
x\rightarrow 0\ ,$ such that $D_{B}$ consists of functions from $D_{b^{+}}$
that have the following asymptotic behavior 
\begin{equation*}
\chi \left( x\right) =B(x^{1/2+\varkappa }+cx^{1/2-\varkappa })+O(x^{3/2}),\
x\rightarrow 0.
\end{equation*}%
As follows from \cite{GitTyV09}, there exists only one s.a. Calogero
Hamiltonian with such a domain. It is $\hat{H}_{2,\lambda }$ with $\lambda
=c>0$,%
\begin{equation}
\hat{H}_{2,c}=\overline{\hat{b}}\hat{b}^{+},\ \ 0<\varkappa <1,\ \varkappa
\neq 1/2\ .  \label{kc.32}
\end{equation}

\subsection{The case c)}

In this case $h\left( x\right) =\mu /x$ ($\alpha \geq -\frac{1}{4},\ \
\varkappa =\sqrt{\alpha +1/4}\geq 0).$ It was demonstrated that $\mu $ is
real and $\mu \in \mathbb{R}_{+}.$ Let us denote%
\begin{align}
& \check{a}_{\mu }=d_{x}-\frac{\mu }{x},\ \ \check{b}_{\mu }=-d_{x}-\frac{%
\mu }{x},\ \ ,  \notag \\
& \check{N}_{\mu }=\check{b}_{\mu }\check{a}_{\mu }=-d_{x}^{2}+\alpha \left(
\mu \right) x^{-2}=\check{H},\ \alpha \left( \mu \right) =\mu ^{2}-\mu , 
\notag \\
& \check{A}_{\mu }=\check{a}_{\mu }\check{b}_{\mu }=-d_{x}^{2}+\alpha
_{1}\left( \mu \right) x^{-2}=\check{H},\ \alpha _{1}\left( \mu \right) =\mu
^{2}+\mu .  \label{kc.34}
\end{align}

We note that in virtue of the obvious relations%
\begin{equation*}
\check{a}_{-\mu }=-\check{b}_{\mu },\ \check{b}_{-\mu }=-\check{a}_{\mu }~,
\end{equation*}%
one can consider only the case $\mu >0.$ Then the differential operations $%
\check{N}_{\mu }$,\ $\mu >0,$ correspond to $\alpha \left( \mu \right) \geq
-1/4,$ whereas $\check{A}_{\mu }$, $\mu >0$ correspond to $\alpha _{1}\left(
\mu \right) >0.$

Let us introduce the initial operators $\hat{a}_{\mu }$, $\hat{b}_{\mu }$, $%
\hat{N}_{\mu }$, and $\hat{A}_{\mu }$defined on the domain $\mathcal{D}%
\left( \mathbb{R}_{+}\right) $ where they act by their differential
operations.

First of all, one can see that the operators $\hat{a}_{\mu }$ and $\hat{b}%
_{\mu }$ are mutually adjoint on $\mathcal{D}\left( \mathbb{R}_{+}\right) $%
\begin{equation*}
\left( \psi ,\hat{a}_{\mu }\xi \right) =\left( \hat{b}_{\mu }\psi ,\xi
\right) ,\;\forall \psi ,\xi \in \mathcal{D}\left( \mathbb{R}_{+}\right) \ .
\end{equation*}%
Second, the initial symmetric operator $\hat{H}$ has the following
representations%
\begin{equation}
\hat{H}=\left\{ 
\begin{array}{l}
\hat{N}_{\mu }=\hat{b}_{\mu }\hat{a}_{\mu }=-d_{x}^{2}+\alpha x^{-2},\ \mu
^{2}-\mu =\alpha \geq -1/4 \\ 
\hat{A}_{\rho }=\hat{a}_{\rho }\hat{b}_{\rho }=-d_{x}^{2}+\alpha x^{-2},\
\rho ^{2}+\rho =\alpha >0%
\end{array}%
\right. .  \label{kc.36a}
\end{equation}

Thus, the initial symmetric operator $\hat{H}$ is nonnegative for $\alpha
\geq -1/4,$%
\begin{equation}
\left( \xi ,\hat{H}\xi \right) =\left( \hat{a}_{\mu }\xi ,\hat{a}_{\mu }\xi
\right) =|\hat{a}_{\mu }\xi |^{2}>0,\;\forall \xi \in \mathcal{D}\left( 
\mathbb{R}_{+}\right) ,\ \alpha \geq -1/4\ .  \label{kc.37}
\end{equation}

Below, we construct closed operators associated to differential expressions $%
\check{a}_{\mu}$ and $\check{b}_{\mu}$.

a) The operator $\hat{a}_{\mu}^{+}$ is defined on the natural domain $D_{%
\check{b}_{\mu}}^{\ast}\left( \mathbb{R}_{+}\right) ,$ i.e., $D_{a_{\mu
}^{+}}=D_{\check{b}_{\mu}}^{\ast}\left( \mathbb{R}_{+}\right) $, where it
acts as $\check{b}_{\mu}$. The functions $\xi\in D_{a_{\mu}^{+}}$ can be
represented as%
\begin{align}
& \xi(x)=cx^{-\mu}+x^{-\mu}\int_{0}^{x}dyy^{\mu}\eta(y),,\;\eta\in L^{2}(%
\mathbb{R}_{+}),  \notag \\
& c=0\mathrm{\ for\ \ }\mu\geq1/2\ .  \label{kc.38}
\end{align}
They have the following asymptotic behavior:%
\begin{align}
& \xi(x)\rightarrow0,\;x\rightarrow\infty,  \label{kc.39} \\
& \xi(x)=cx^{-\mu}+O(x^{1/2}),\;x\rightarrow0.  \notag
\end{align}

b) The operator $\hat{b}_{\mu}^{+}$ is defined on the natural domain $D_{%
\check{a}_{\mu}}^{\ast}\left( \mathbb{R}_{+}\right) ,$ i.e., $D_{b_{\mu
}^{+}}=D_{\check{a}_{\mu}}^{\ast}\left( \mathbb{R}_{+}\right) $, where it
acts as $\check{a}_{\mu}$. The functions $\xi\in D_{b_{\mu}^{+}}$ can be
represented as:

For $\mu>1/2:$%
\begin{align}
\xi(x) & =x^{\mu}\int_{x}^{\infty}dyy^{-\mu}\eta(y),\;\eta\in L^{2}(\mathbb{R%
}_{+}),  \notag \\
\xi(x) & \rightarrow0,\;x\rightarrow\infty;\ \
\xi(x)=O(x^{1/2}),\;x\rightarrow0.  \label{kc.40}
\end{align}

For $\mu=1/2:$%
\begin{align}
\xi(x) & =cx^{1/2}+x^{1/2}\int_{x_{0}}^{x}dyy^{-1/2}\eta(y),\;\eta\in L^{2}(%
\mathbb{R}_{+}),  \notag \\
\xi(x) & \rightarrow0,\;x\rightarrow\infty;\ \ \xi(x)=O(x^{1/2}\ln
x),\;x\rightarrow0.  \label{kc.41}
\end{align}

For $\mu<1/2:$%
\begin{align}
\xi(x) & =cx^{\mu}+x^{\mu}\int_{0}^{x}dyy^{-\mu}\eta(y),\;\eta\in L^{2}(%
\mathbb{R}_{+}),  \notag \\
\xi(x) & \rightarrow0,\;x\rightarrow\infty;\ \
\xi(x)=cx^{\mu}+O(x^{1/2}),\;x\rightarrow0.  \label{kc.43}
\end{align}

c) The operator $\overline{\hat{a}_{\mu}}$ is defined on the domain%
\begin{equation}
D_{\bar{a}_{\mu}}=\left\{ 
\begin{array}{l}
D_{b_{\mu}^{+}},\;\mu\geq1/2, \\ 
\left\{ \xi:\xi\in D_{b_{\mu}^{+}},\;\xi(x)=O(x^{1/2}),\,x\rightarrow
0,\right\} ,\;\mu<1/2,%
\end{array}
\right.  \label{kc.44}
\end{equation}
where it acts as $\check{a}_{\mu}.$

d) The operator $\overline{\hat{b}_{\mu}}$ is defined on the domain 
\begin{equation}
D_{b_{\mu}}=\left\{ 
\begin{array}{l}
D_{a_{\mu}^{+}},\;\mu\geq1/2, \\ 
\left\{ \xi:\xi\in D_{a_{\mu}^{+}},\;\xi(x)=O(x^{1/2}),\;x\rightarrow
0\right\} ,\;\mu<1/2,%
\end{array}
\right.  \label{kc.45}
\end{equation}
where it acts as $\check{b}_{\mu}.$

Note that $\overline{\hat{b}_{\mu}}=\hat{a}_{\mu}^{+}$ and $\hat{b}%
_{\mu}^{+}=\overline{\hat{a}_{\mu}}$ for $\mu\geq1/2$; but this is not true
for $\mu<1/2$.

We see that there are the following relations between the domains of the
closed operators $\overline{\hat{a}_{\mu }}$, $\overline{\hat{b}_{\mu }}$, $%
\hat{a}_{\mu }^{+}$, and $\hat{b}_{\mu }^{+}$:%
\begin{align}
D_{a_{\mu }^{+}}& =\left\{ 
\begin{array}{l}
D_{\bar{b}_{\mu }}\ ,\;\mu \geq 1/2, \\ 
D_{\bar{b}_{\mu }}+A\chi _{0},\ \chi _{0}=x^{-\mu }\psi _{0}(x),\ 0<\mu <1/2,%
\end{array}%
\right.  \notag \\
D_{b_{\mu }^{+}}& =\left\{ 
\begin{array}{l}
D_{\bar{a}_{\mu }}\ ,\;\mu \geq 1/2, \\ 
D_{\bar{a}_{\mu }}+B\psi _{0},\ \psi _{0}=x^{\mu }\zeta (x),\ 0<\mu <1/2.%
\end{array}%
\right.  \label{kc.46}
\end{align}%
Here $A,B$ are arbitrary constants.

Thus, in the case under consideration, we have demonstrated that a mutually
conjugated by Lagrange pair of differential operations $\check{a}_{\mu }$
and $\check{b}_{\mu }$ generates two mutually conjugated closed operator
pairs $\overline{\hat{b}_{\mu }}$ and $\hat{b}_{\mu }^{+}$ and $\overline{%
\hat{a}_{\mu }}$ and $\hat{a}_{\mu }^{+}.$ On their domains the operators $%
\overline{\hat{b}_{\mu }}$ and $\hat{a}_{\mu }^{+}$ act as $\check{b}_{\mu
}, $ and the operators $\hat{b}_{\mu }^{+}$ and $\overline{\hat{a}_{\mu }}$
act as $\check{a}_{\mu }.$ At the same time the above results allows one to
assert that the are no other conjugated closed pairs that obey natural
properties similar (\ref{KC.8c}).

Now we are in position to construct oscillator representations for some s.a.
Calogero Hamiltonians in the case under consideration.

Let us introduce the operators%
\begin{align}
\hat{N}_{1}\left( \mu \right) & =\hat{a}_{\mu }^{+}\overline{\hat{a}_{\mu }}%
,\;\hat{N}_{2}\left( \rho \right) =\hat{b}_{\rho }^{+}\overline{\hat{b}%
_{\rho }};\   \notag \\
\hat{A}_{1}\left( \rho \right) & =\overline{\hat{a}_{\rho }}\hat{a}_{\rho
}^{+},\;\hat{A}_{2}\left( \mu \right) =\overline{\hat{b}_{\mu }}\hat{b}_{\mu
}^{+}\ .  \label{kc.51}
\end{align}%
According to the Akhiezer Glazman Theorem, all these operators are
nonnegative s.a. operators. According to (\ref{kc.36a}) all the operators
are s.a. extensions of the initial symmetric operator $\hat{H}$ with some $%
\alpha $, and therefore are some s.a. Calogero Hamiltonians. Below, we
identify them with such Hamiltonians using results of \cite{GitTyV09}.

a) Consider the operator $\hat{N}_{1}\left( \mu\right) =\hat{a}_{\mu}^{+}%
\hat{a}_{\mu}$, $\mu^{2}-\mu=\alpha$.

Let $\mu \geq 3/2$, $\alpha \geq 3/4.$ In this case, there exists only one
s.a. Calogero Hamiltonian $\hat{H}_{1},$ such that%
\begin{equation}
\hat{H}_{1}=\hat{a}_{\mu }^{+}\overline{\hat{a}_{\mu }},\ \ \mu \geq 3/2,\
\alpha \geq 3/4.  \label{kc.52}
\end{equation}

Let $3/2>\mu >1/2$, $3/4>\alpha >-1/4.$ In this case the condition $\xi \in
D_{N_{1}\left( \mu \right) }$ implies that $\xi \in D_{b_{\mu }^{+}}$. In
turn, the latter condition implies that $\xi (x)$ tend to zero not weaker
than $x^{1/2}$ as $x\rightarrow 0$. For $3/4>\alpha >-1/4$, the only
functions from $D_{H_{2,0}}$ (the domain of s.a. Calogero Hamiltonian $\hat{H%
}_{2,0}$) have such an asymptotic behavior. Therefore, we can conclude that%
\begin{equation}
\hat{H}_{2,0}=\hat{a}_{\mu }^{+}\overline{\hat{a}_{\mu }},\ 3/2>\mu >1/2,\
3/4>\alpha >-1/4.  \label{kc.53}
\end{equation}

Let $\mu =1/2$, $\alpha =-1/4.$ In this case the condition $\xi \in
D_{N_{1}\left( 1/2\right) }$ implies that $\xi \in
D_{a_{1/2}}=D_{b_{1/2}^{+}}$. If we represent $\xi $ as (\ref{kc.41}), then $%
\hat{a}_{1/2}\xi =\eta $, which implies that $\eta \in D_{a_{1/2}^{+}}$ and $%
\eta (x)=O(x^{1/2})$ as $x\rightarrow 0$. From the same representation (\ref%
{kc.41}), using obtained asymptotic behavior of $\eta (x)$, we find $\xi
(x)=O(x^{1/2})$ as $x\rightarrow 0$. Thus,%
\begin{equation}
\hat{H}_{3,\infty }=\hat{a}_{1/2}^{+}\overline{\hat{a}_{1/2}},\ \ \alpha
=-1/4\ .  \label{kc.54}
\end{equation}

Let $1/2>\mu >0$, $-1/4<\alpha <0$. In this case, $\xi \in D_{N_{1}\left(
\mu \right) }$ implies that $\xi \in D_{a_{\mu }}$, and therefore $\xi (x)$
tend to zero not weaker than $x^{1/2}$ as $x\rightarrow 0$. For $-1/4<\alpha
<0$ the only functions from $D_{H_{2,\infty }}$ (the domain of s.a. Calogero
Hamiltonian $\hat{H}_{2,\infty }$) have such an asymptotic behavior.
Therefore, we can conclude that%
\begin{equation}
\hat{H}_{2,\infty }=\hat{a}_{\mu }^{+}\hat{a}_{\mu },\ \ 1/2>\mu >0,\
-1/4<\alpha <0.  \label{kc.55}
\end{equation}

b) Consider the operator $\hat{A}_{2}\left( \mu\right) =\overline{\hat {b}%
_{\mu}}\hat{b}_{\mu}^{+}$, $\mu^{2}-\mu=\alpha$.

Let $\mu \geq 1/2$, $\alpha \geq -1/4.$ In this case,%
\begin{align}
& \hat{H}_{1}=\overline{\hat{b}_{\mu }}\hat{b}_{\mu }^{+},\ \mu \geq 3/2,\
\alpha \geq 3/4,  \notag \\
& \hat{H}_{2,0}=\overline{\hat{b}_{\mu }}\hat{b}_{\mu }^{+},\ 3/2>\mu >1/2,\
3/4>\alpha >-1/4,  \notag \\
& \hat{H}_{3,\infty }=\overline{\hat{b}_{1/2}}\hat{b}_{1/2}^{+},\ \mu =1/2,\
\alpha =-1/4\ .  \label{kc.56}
\end{align}

Let $1/2>\mu>0$, $-1/4<\alpha<0.$ In this case $\xi\in D_{A_{2}\left(
\mu\right) }$ implies $\xi\in D_{b_{\mu}^{+}}$. Let us represent $\xi$ in
the form (\ref{kc.43}), $\hat{b}_{\mu}^{+}\xi=\eta$, i.e., $\eta\in D_{\hat {%
b}_{\mu}}$. This implies that $\eta(x)=O(x^{1/2})$ as $x\rightarrow0$. From
the same representation (\ref{kc.43}), using the obtained asymptotic
behavior of $\eta(x)$ we find that $\xi(x)=cx^{\mu}+O(x^{3/2})\;$as $%
x\rightarrow0$. For $-1/4<\alpha<0$ the only functions from $D_{H_{2,0}}$
(the domain of s.a. Calogero Hamiltonian $\hat{H}_{2,0}$) have such an
asymptotic behavior. Therefore, we can conclude that%
\begin{equation}
\hat{H}_{2,0}=\overline{\hat{b}_{\mu}}\hat{b}_{\mu}^{+},\ 1/2>\mu >0,\ \
-1/4<\alpha<0.  \label{kc.57}
\end{equation}

We note that we $\hat{A}_{2}\left( \mu\right) \neq\hat{N}_{1}\left(
\mu\right) $ in the case under consideration.

c) Consider the operator $\hat{A}_{1}\left( \rho\right) =\overline{\hat {a}%
_{\rho}}\hat{a}_{\rho}^{+},\ \rho^{2}+\rho=\alpha$.

Let $\rho \geq 1/2,$ $\alpha \geq 3/4.$ In this case, we have%
\begin{equation}
\hat{H}_{1}=\hat{a}_{\rho }\overline{\hat{a}_{\rho }^{+}},\ \rho \geq 1/2,\
\alpha \geq 3/4.  \label{kc.58}
\end{equation}

Let $0<\rho <1/2$, $0<\alpha <3/4.$ In this case $\xi \in D_{A_{1}\left(
\rho \right) }$ implies $\xi \in D_{a_{\rho }^{+}}$\thinspace . Let us
represent $\xi $ in the form (\ref{kc.38}), $\hat{a}_{\rho }^{+}\xi =\eta $,
i.e., $\eta \in D_{\hat{a}_{\rho }}$. This implies that $\eta (x)=O(x^{1/2})$
as $x\rightarrow 0$. From the same representation (\ref{kc.38}), using the
obtained asymptotic behavior of $\eta (x)$, we find that $\xi (x)=cx^{-\rho
}+O(x^{3/2}),\;x\rightarrow 0$. For $0<\alpha <3/4$ the only functions from $%
D_{H_{2,\infty }}$ (the domain of s.a. Calogero Hamiltonian $\hat{H}%
_{2,\infty }$) have such an asymptotic behavior. Therefore, we can conclude
that%
\begin{equation}
\hat{H}_{2,\infty }=\hat{a}_{\rho }\overline{\hat{a}_{\rho }^{+}},\ 0<\rho
<1/2,\ 0<\alpha <3/4.  \label{kc.59}
\end{equation}

d) Consider the operator $\hat{N}_{2}\left( \rho\right) =\hat{b}_{\rho}^{+}%
\overline{\hat{b}_{\rho}}$, $\rho^{2}+\rho=\alpha.$

Let $\rho\geq1/2$, $\alpha\geq3/4.$ In this case, we have%
\begin{equation}
\hat{H}_{1}=\hat{b}_{\rho}^{+}\overline{\hat{b}_{\rho}},\ \rho\geq 1/2,\
\alpha\geq3/4.  \label{kc.60}
\end{equation}

Let $1/2>\rho >0$, $3/4>\alpha >0.$ In this case, $\xi \in D_{N_{2}\left(
\rho \right) }$ implies $\xi \in D_{\bar{b}_{\rho }}$, and therefore $\xi
(x) $ tend to zero not weaker than $x^{1/2}$ as $x\rightarrow 0$. For $%
3/4>\alpha >0$ the only functions from $D_{H_{2,0}}$ (the domain of s.a.
Calogero Hamiltonian $\hat{H}_{2,0}$) have such an asymptotic behavior.
Therefore, we can conclude that%
\begin{equation}
\hat{H}_{2,0}=\hat{b}_{\rho }^{+}\overline{\hat{b}_{\rho }},\ 1/2>\rho >0,\
\ 3/4>\alpha >0.  \label{kc.61}
\end{equation}

We stress that $\hat{N}_{2}\left( \rho \right) \neq \hat{A}_{1}\left( \rho
\right) $ in the case under consideration.

In the conclusion, we note that all nonnegative s.a. Calogero Hamiltonians $%
\hat{H}_{1}$, $\hat{H}_{2,\lambda },\ \lambda \geq 0$, $\hat{H}_{2,\infty }$%
, and $\hat{H}_{3,\infty }$ from \cite{GitTyV09} were represented above in
the oscillator form. As to s.a. extensions of a closed symmetric nonnegative
operator with finite deficiency indices $(m,m)$, it is known that the
negative part of the spectrum for each of its s.a. extensions can only
consist of negative eigenvalues, the sum of whose multiplicities does not
exceed $m$, and there exist s.a. extensions with a nonnegative spectrum
(with the preservation of the infimum), see \cite{AkhGl81,Naima69}. This
general property of nonnegative operators explains the remarkable fact that
for $3/4>\alpha \geq -1/4$ the number of negative levels of each s.a.
Hamiltonian does not exceed unity, because in this case $m=1$, and the
negative spectrum is absent for $\alpha \geq 3/4$, since in this case $m=0$.
These properties explain also the fact that only $\hat{H}$ with $\alpha \geq
-1/4$ are represented as a product of two mutually-conjugated operators. For 
$\alpha <-1/4$ the number of negative eigenvalues is infinite.

\subparagraph{Acknowledgement}

Gitman is grateful to the Brazilian foundations FAPESP and CNPq for
permanent support; Tyutin thanks FAPESP and RFBR, grant 08-02-01118; Tyutin
and Voronov thank LSS-1615.2008.2 for partial support.

\end{document}